\documentclass[aip, apl, longbibliography, superscriptaddress, amssymb, reprint]{revtex4-1}
\usepackage{siunitx,graphicx,dcolumn,bm,physics}
\usepackage{graphicx,physics}
\usepackage{amsmath}
\usepackage{multirow}

\newcommand{\VSD}{V_\text{SD}}
\newcommand{\VS}{V_\text{S}}
\newcommand{\VG}{V_\text{G}}
\newcommand{\VM}{V_\text{M}}
\newcommand{\Vout}{V_\text{out}}
\newcommand{\VN}{V_\text{N}}

\newcommand{\GammaV}{\Gamma_V}

\newcommand{\Pout}{\ensuremath{P_\mathrm{out} }}
\newcommand{\IDC}{I_\text{DC}}
\newcommand{\Ztrans}{Z_\text{trans}}

\newcommand{\fE}{f_\text{E}}
\newcommand{\fM}{f_\text{M}}
\newcommand{\chiM}{\chi_\text{M}}
\newcommand{\QM}{Q_\text{M}}

\newcommand{\CS}{C_\text{S}}
\newcommand{\CNT}{C_2}

\newcommand{\NW}{N_\text{W}}

\begin{document}

\title{Measuring carbon nanotube vibrations using a single-electron transistor as a fast linear amplifier}

\author{Yutian~Wen}
\affiliation{Department of Materials, University of Oxford, Parks Road, Oxford OX1 3PH, United Kingdom}

\author{N.~Ares}
\affiliation{Department of Materials, University of Oxford, Parks Road, Oxford OX1 3PH, United Kingdom}

\author{T.~Pei}
\affiliation{Department of Materials, University of Oxford, Parks Road, Oxford OX1 3PH, United Kingdom}

\author{G.A.D.~Briggs}
\affiliation{Department of Materials, University of Oxford, Parks Road, Oxford OX1 3PH, United Kingdom}

\author{E.A.~Laird}
\email{e.a.laird@lancaster.ac.uk}
\affiliation{Department of Physics, Lancaster University, Lancaster, LA1 4YB, United Kingdom}
\affiliation{Department of Materials, University of Oxford, Parks Road, Oxford OX1 3PH, United Kingdom}

\date{\today}
\begin{abstract}
  We demonstrate sensitive and fast electrical measurements of a carbon nanotube mechanical resonator. The nanotube is configured as a single-electron transistor, whose conductance is a sensitive transducer for its own displacement. Using an impedance-matching circuit followed by a cryogenic amplifier, the vibrations can be monitored at radio frequency. The sensitivity of this continuous displacement measurement approaches within a factor 470 of the standard quantum limit.
\end{abstract}
\maketitle

Suspended carbon nanotubes are mechanical resonators~\cite{Sazonova2004} with low mass, high compliance, and high quality factor~\cite{Huttel2009,Moser2014}, which makes them sensitive electromechanical detectors for tiny forces~\cite{Moser2013} and masses~\cite{Chiu2008,Lassagne2008,Chaste2012}.
The time-averaged current through a vibrating nanotube probes electron-phonon coupling~\cite{Leturcq2009, Steele2009,Lassagne2009,Benyamini2014}, non-linear dissipation~\cite{Eichler2011}, and mechanical mode mixing~\cite{Castellanos-Gomez2012} on the nanoscale.
Time-resolved measurements go further, allowing for the study of transient effects such as spin switching~\cite{Ganzhorn2013,Willick2014}, mechanical dephasing~\cite{Schneider2014a}, or even force-detected magnetic resonance~\cite{Poggio2018}.
Although the low mass favors large electromechanical coupling, the large electrical impedance of nanotube devices makes it difficult to amplify the current signal with high sensitivity and bandwidth, especially since low temperatures are needed to suppress thermal noise.

One approach is to downconvert the electromechanical signal to a frequency within the bandwidth of a cryogenic current amplifier, using either two-source mixing or the non-linear conductance of the nanotube itself~\cite{Meerwaldt2013, Willick2017a}. For fast measurements, parasitic capacitance must be minimised by placing the amplifier close to the resonator.
The resulting heat load has usually prevented operation below 1~K~\cite{Schneider2014a, Willick2017a}, although recently such a setup achieved high sensitivity at millikelvin temperatures and with a bandwidth of 87~kHz~\cite{deBonis2018}.
A second approach, with higher bandwidth, is to detect the changing capacitance between the vibrating nanotube and a pickup antenna~\cite{Ares2016a}. However, the small size of the pickup antenna means that a large electric field is needed to generate an appreciable signal.
A third approach, used for other kinds of nanoscale resonator~\cite{Xu2010,Mathew2015}, is to connect the resonator's output directly to a fast amplifier matched to the cable impedance (typically $50~\Omega$). However, the electrical divider formed between the large impedance of the device and the small impedance of the cable degrades the signal.

Here, we demonstrate a circuit that combines sensitivity with high speed by monitoring the electromechanical signal directly, while requiring only a DC bias voltage.
The circuit exploits a single-electron transistor (SET) defined within the nanotube as the initial stage of displacement amplification~\cite{Knobel2003,LaHaye2004,Huttel2009,Steele2009,Lassagne2009,Laird2012,Benyamini2014,Pashkin2010}.
The SET output current, which depends linearly on displacement, is monitored directly at radio frequency using a low-noise cryogenic radio-frequency (RF) amplifier with MHz bandwidth.
To improve the coupling between the SET and the cryogenic amplifier, we use an impedance-matching stage based on a tuneable RF tank circuit~\cite{Ares2016}.
We characterize the displacement imprecision achieved in this setup and show that it approaches within a factor $470$ of the limit set by quantum mechanics.
This technique combines the speed allowed by RF readout with the high sensitivity of an SET amplifier integrated into the moving nanotube.

\begin{figure}
\includegraphics[width=8.5cm]{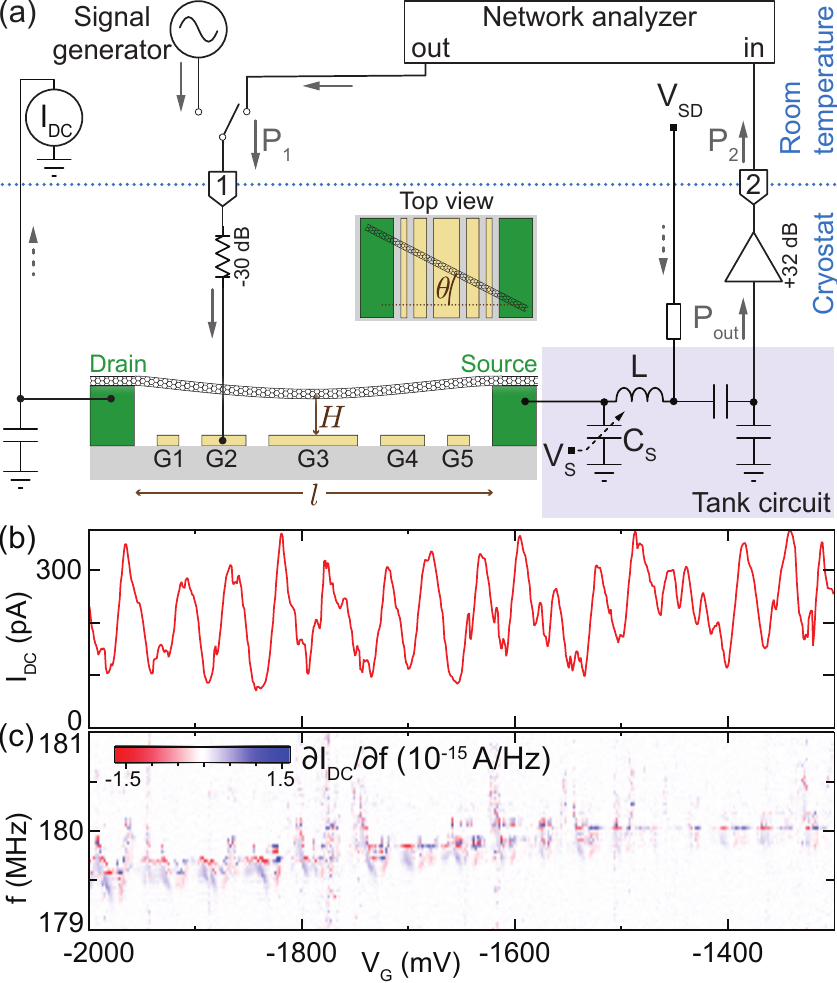}
\caption{\label{fig:1} (a) Device and measurement setup. The vibrating nanotube is suspended between source and drain electrodes and over five gate electrodes that define a quantum dot potential. Gate 2 is connected to an attenuated high-frequency line for mechanical actuation.
The device is biased by voltage $\VSD$ and measured both via DC transport and via a tuneable RF circuit (see text). The DC current path is marked by dashed arrows, the RF path by solid arrows.
The length $l$ and suspended height $H$ of the nanotube are indicated in the main diagram, and an inset marks the alignment angle $\theta$.
(b) Coulomb blockade peaks measured in DC current as a function of DC voltage $\VG$ on gate 2, with $\VSD=5~$mV.
(c) Characterization of the mechanical resonance in transport. With RF power applied at frequency $f$, the derivative $\partial \IDC/ \partial f$ shows the resonance as a weakly gate-dependent feature at $f \sim 180$~MHz. The RF power at port 1 was $P_1 = -35$~dBm, and the bias was $\VSD=5$~mV. In panels (b) and (c), the other gate voltages were held at positive voltages between 0 and 300~mV.}
\end{figure}

The experimental setup is shown in Fig.~\ref{fig:1}(a).
The resonator is fabricated by stamping a carbon nanotube across lithographically patterned Cr/Au contact electrodes, giving a suspended length of $l=800$~nm~\cite{Wu2010,Pei2012,Mavalankar2016}.
Beneath the nanotube, five predefined Cr/Au finger gates, labelled G1-5, allow tuning of the electrical potential along the nanotube.
This device is bonded to a printed circuit board and loaded into a 25~mK dilution refrigerator.
For electromechanical excitation, gate G2 is connected via a bias tee to an RF drive line (port 1 in Fig.~\ref{fig:1}(a)), and is driven either by a RF signal generator or using the output of a network analyzer.

With the drain electrode held at ground, a source-drain bias $V_\mathrm{SD}$, applied to the source electrode, drives a current through the nanotube which can be measured both at DC and at RF.
The DC current $I_\mathrm{DC}$ is monitored using a room-temperature current amplifier connected to the drain.
The RF response is measured using a cryogenic amplifier anchored at 5~K inside the cryostat.
A resonant tank circuit (Fig.~\ref{fig:1}(a) right), constructed using chip inductors and capacitors on the circuit board and bonded to the source electrode, is interposed between the device and the amplifier to transduce the RF current through the nanotube to an output voltage $V_\mathrm{out}$, which is fed to the cryogenic amplifier~\cite{Ares2016a}. This amplified signal is fed via port~2 of the cryostat into the network analyzer.
The tank circuit incorporates a variable capacitor $C_\mathrm{S}$, controlled with a tuning voltage $V_\mathrm{S}$, which adjusts the electrical resonance frequency~\cite{Ares2016}.

To allow sensitive electromechanical measurements, the suspended nanotube is operated as an SET.
This SET is formed between the Schottky barriers along the nanotube and tuned using DC voltages applied to the finger gates~\cite{Mason2004, Laird2015}.
Figure~\ref{fig:1}(b) plots the DC current $I_\mathrm{DC}$ through the nanotube as a function of the DC voltage $\VG$ applied to gate G2. The pattern of current peaks indicates Coulomb blockade, confirming SET behavior.

We first identify the mechanical resonance using DC transport. With an RF drive at frequency $f$ applied to gate~G2,
Fig.~\ref{fig:1}(c) shows the derivative $\partial \IDC/\partial f$ as a function of $f$ and $\VG$.
The current has a sharp peak or dip when $f$ matches the mechanical resonance frequency $\fM$, leading to a feature in the derivative~\cite{Sazonova2004, Huttel2009} appearing around $\fM\approx 180$~MHz.
The resonance frequency increases with increasing $\VG$ as the mechanical tension changes.
It is also modulated by Coulomb blockade, because electron tunnelling modifies the effective spring constant~\cite{Steele2009, Lassagne2009}.

\begin{figure}
	\includegraphics[width=3.37in]{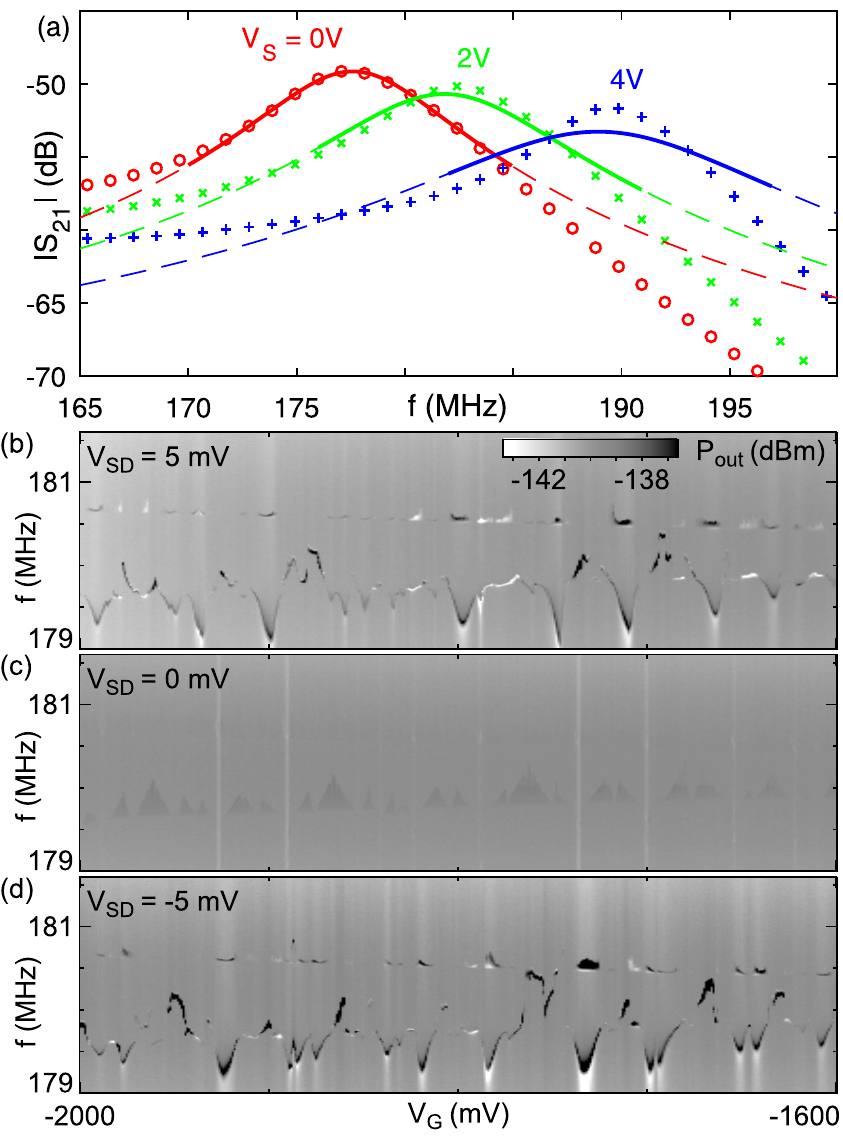}
	\centering
	\caption{\label{fig:2} Characterization of the tank circuit and the mechanical resonance (a) Electrical transmission from port~1 to port~2, for different settings of the tuning voltage $\VS$ and with $\VSD=0$~mV. Symbols: Data. The main electrical resonance appears as a tuneable transmission peak in the range $178-205$~MHz.
	Curves: Fits to an electrical model of the tank circuit (see supplementary material). The solid section of each curve indicates the fitting range. (b-d) Transmitted signal, converted to tank circuit output power $\Pout$, as a function of frequency and gate voltage for different settings of $\VSD$. Here $\VS=0$~V and drive power is $P_1 = -70$~dBm. The main mechanical resonance is evident as a sharp feature in (b) and (d), whose frequency varies with gate voltage because of Coulomb blockade. A weaker resonance appears at slightly higher frequency. This may be a mode vibrating approximately in the plane of the sample, i.e. orthogonal to the stronger resonance, and is not studied further.}
\end{figure}

We now turn to RF measurements. To optimise the sensitivity, we first tune the tank circuit's electrical resonance frequency $\fE$ to a value near $\fM$. This is inferred from the transmission $S_{21}$ from port~1 to port~2, measured using a network analyzer at different settings of the varactor tuning voltage $\VS$ (Fig.~\ref{fig:2}(a)). The main tank circuit resonance is evident as a broad transmission peak, whose frequency increases as $\CS$ is tuned towards lower values. By fitting these traces, we are able to extract the tank circuit parameters (see supplementary material). In the rest of this paper, we fix $\VS=0$~V, giving optimum sensitivity around $\fE\approx 178$~MHz and a detection bandwidth of $\approx 8$~MHz, set by the electrical quality factor.

The mechanical signal appears as a sharp resonance superimposed on the electrical transmission peak when a source-drain bias $\VSD$ is applied.
This is evident in Figs.~\ref{fig:2}(b-d), which show the power $\Pout$ transmitted from the output of the tank circuit, plotted as a function of frequency and gate voltage.
This signal arises because the motion of the nanotube relative to the gates modulates the SET conductance, leading to a current oscillating at the vibration frequency.
This current is transduced to an RF output voltage $\Vout$, with  $\Pout =\Vout^2/Z_0$, where $Z_0=50~\Omega$ is the line impedance.
As expected from this mechanism, the signal appears for both positive and negative bias (Fig.~\ref{fig:2}(b,d)), but nearly vanishes at zero bias (Fig.~\ref{fig:2}(c)).
The fact that the signal is larger with a bias applied confirms that it arises mainly from the SET conductance.
The weak signal remaining at zero bias indicates a small capacitive contribution~\cite{Ares2016a}.
We interpret this mode as the fundamental out-of-plane flexural mode.

\begin{figure}
	\includegraphics[width=3.37in]{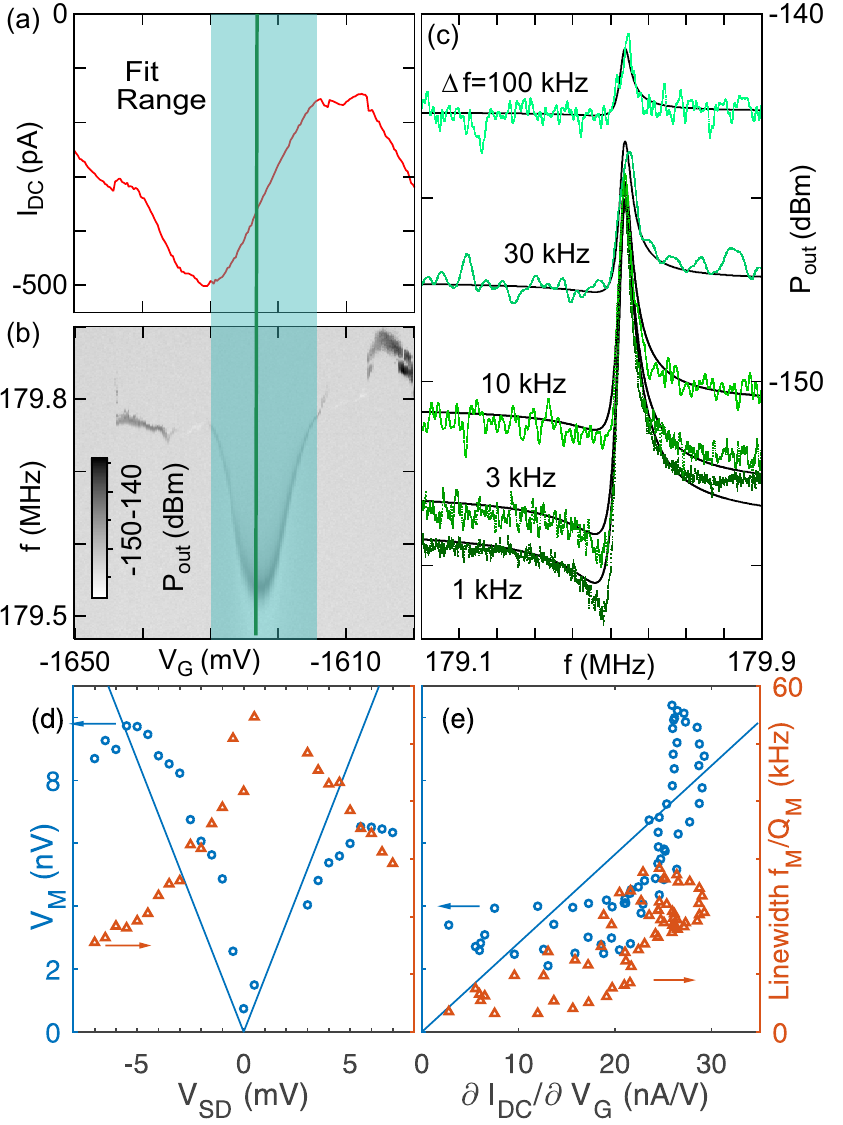}
	\centering
	\caption{\label{fig:3} Measuring the displacement sensitivity
		(a) DC current as a function of gate voltage in the region of one Coulomb blockade peak, with $\VSD=-5$~mV.
		(b) Mechanical resonance measured in RF power simultaneously with (a).
		(c) Transmission as a function of frequency, measured with gate excitation voltage $\delta \VG = 1.2~\mu$V and for different settings of the network analyzer's resolution bandwidth $\Delta f$.
		Points: data; Curves: fits to Eq.~\eqref{eq:Pout}. Measurements are taken on the flank of the Coulomb blockade peak (vertical line in (a-b)).
		(d) Mechanical signal(circles, left axis) and linewidth (triangles, right axis) as functions of bias. Line: Fit of the form $\VM \propto |\VSD|$.
		(e) The same quantities as functions of  transconductance. Line: Fit of the form $\VM \propto \partial \IDC/ \partial \VG$.
	}
\end{figure}

We now characterize the displacement sensitivity. We do this by measuring the signal and the noise of the output voltage when the nanotube is driven to a fixed displacement amplitude, by adding a known oscillating gate voltage $\delta \VG$ to $\VG$.
This measurement is performed with the gate voltage tuned to the flank of a Coulomb blockade peak (Fig.~\ref{fig:3}(a)).
As expected, this gate voltage leads to a strong mechanical signal (Fig.~\ref{fig:3}(b)).

Both signal and noise are extracted from Fig.~\ref{fig:3}(c), which shows the power spectrum $\Pout$, measured with root-mean-square driving amplitude $\delta \VG=1.3~\mu$V and for different settings of the network analyzer's resolution bandwidth $\Delta f$. The mechanical resonance is evident at frequency $\fM \approx 179.55$~MHz. Whereas a purely mechanical response would lead to a symmetric Lorentzian peak, the observed peak is asymmetric, indicating a small contribution from stray electronic transmission in the sample holder\cite{Supplementary}. To extract the mechanical signal strength, we fit these traces with the following function, which takes account of the resonant mechanical response, a non-resonant electrical background, and broadband detection noise (see supplementary material):
\begin{equation}
P_\mathrm{out} (f) = \frac{A^2}{Z_0} \left| B e^{i\phi_B} + \frac{\fM^2/\QM}{\fM^2 - f^2 + i \frac{f \fM}{\QM}}\right|^2 \delta \VG^2+ \frac{S_{VV} \Delta f}{Z_0},
\label{eq:Pout}
\end{equation}
where $Z_0=50~\Omega$ is the line impedance, $\QM$ is the mechanical quality factor, $S_{VV}$ is the one-sided spectral density of the system voltage noise, $\Delta f$ is the resolution bandwidth, and $A$, $B$, and $\phi_B$ are constants.
By fitting Fig.~\ref{fig:3}(c) to Eq.~\eqref{eq:Pout}, using $\fM$, $\QM$, $S_{VV}$, $A$, $B$, and $\phi_B$ as global fit parameters, we find a voltage sensitivity $\sqrt{S_{VV}}=50.3 \pm 0.5~\mathrm{pV}/\sqrt{\mathrm{Hz}}$ and a quality factor $\QM=6000 \pm 1500$.
From Fig~\ref{fig:3}(b), it is clear that the linewidth varies strongly with gate voltage across the Coulomb peak, which implies that the quality factor is limited by dissipation due to inelastic electron tunneling~\cite{Steele2009,Lassagne2009}.
The fitting function does not take account of possible mechanical non-linearity, which is justified by the good quality of the fits in Fig.~\ref{fig:3}(c).
We have also confirmed that this this drive amplitude is below the observable onset of Duffing distortion of the lineshape, and is also below the onset of power-broadening~\cite{Meerwaldt2012a}.

On resonance, the purely electromechanical part of the signal in Fig.~\ref{fig:3}(c) is:
\begin{align}
V_\mathrm{M} 	&= A\, \delta \VG	\\
				&= 8.79 \pm 0.08~\mathrm{nV}.	\label{eq:VM}
\end{align}
To quantify the sensitivity, we must relate this output voltage amplitude to the corresponding displacement $u$. The displacement depends on the driving force $\delta F$ via:
\begin{align}
 u &= |\chiM(f)|\, \delta F	\\
   &= |\chiM(f)|\, \VG \frac{ \partial \CNT}{\partial u}\, \delta \VG.
\label{eqn:input}
\end{align}
where where $\CNT$ is the capacitance between the gate and the nanotube. Here $\chiM(f) \equiv \frac{1}{4\pi^2m}[\fM^2-f^2 + i \frac{f \fM}{\QM}]^{-1}$ is the mechanical susceptibility~\cite{Aspelmeyer2014} at driving frequency $f$, and $m$ is the mass. The force is $\delta F=\VG \frac{ \partial \CNT}{\partial u}\, \delta \VG$.
We must therefore estimate the parameters in Eq.~(\ref{eqn:input}).
This is complicated by the uncertainty in the nanotube's mass, its diameter, and in the suspended height above the gate, which affects the capacitance derivative. The full estimation procedure is described in the supplementary material. Using the known gate capacitance $\CNT\approx 3.3\pm 0.6$~aF, extracted from a Coulomb blockade measurement similar to Fig.~\ref{fig:1}(b), we use finite-element electrostatic simulation to deduce that the height is $H=18_{-13.5}^{+92}$~nm. Combining this value with the nanotube diameter $D=4.5 \pm 1.5$~nm, estimated from transmission electron microscopy, leads to a capacitance derivative $\partial \CNT/{\partial u} = 49.4^{+271}_{-40.5}$~pF~m$^{-1}$. The error range is dominated by the uncertainty in the suspended height and suspension angle~$\theta$.
The nanotube's mass is $m=21.3^{+81}_{-15.5}$~ag, with the uncertainty arising from the unknown number of walls and from the suspension angle~$\theta$.

Substituting these quantities into Eq.~\eqref{eqn:input} allows us to calculate that in Fig.~\ref{fig:3}(c) the mechanical amplitude is
$u=25.2^{+811}_{-24.3}~\mathrm{pm}$.
The proportionality constant between the signal voltage (Eq.~\eqref{eq:VM}) and the corresponding displacement is therefore known, finally leading to a displacement sensitivity, defined as the square root of the measurement imprecision:
\begin{align}
\sqrt{S_{uu}} 	&= \frac{u}{V_\mathrm{M}} \sqrt{S_{VV}} \\
	&= 144^{+4700}_{-139}~\mathrm{fm/\sqrt{Hz}}.
\label{eq:Su}
\end{align}
The large range of this estimate reflects the combination of uncertainties entering Eq.~\eqref{eqn:input}.

In this model, the electromechanical signal $\VM$ should increase with $|\VSD|$, because a larger bias leads to a larger mechanically modulated current.
This is tested in Fig.~\ref{fig:3}(d), which plots $\VM$ against $\VSD$.
As expected, the signal is approximately proportional to bias. At high bias, the signal falls below the trend because the SET Coulomb peaks become less sharp.
The electromechanical signal should also be proportional to the SET transconductance $\partial \IDC / \partial \VG$.
This is tested in Fig.~\ref{fig:3}(e), which plots $\VM$ against  transconductance at fixed bias.
The data show approximate proportionality, again confirming that the signal arises mainly from conductance through the SET.

We now compare the experimental displacement sensitivity with what would be achieved by a quantum-limited detector~\cite{LaHaye2004}.
In a continuous phase-preserving measurement, minimum uncertainty requires the imprecision noise to be equal to the noise generated by backaction.
In this ideal case, the displacement sensitivity obeys the standard quantum limit (SQL)~\cite{Regal2008, Clerk2010}:
\begin{align}
\sqrt{S_{uu} (\mathrm{SQL})}	&= \sqrt{\frac{\hbar \QM}{4\pi^2m\fM^2}}	\\
				&= 4.8^{+5.5}_{-2.9}~\mathrm{fm/\sqrt{Hz}}.
\label{eq:SuSQL}
\end{align}
Comparing Eq.~\eqref{eq:SuSQL} with Eq.~\eqref{eq:Su} gives the normalised sensitivity:
\begin{equation}
\frac{\sqrt{S_{uu}}}{\sqrt{S_{uu}(\mathrm{SQL})}} = 30^{+440}_{-27}
\label{eq:SQLratio}
\end{equation}
meaning that this measurement is within a factor 470 of the SQL.
A larger mass than estimated, for example because of surface contamination, would imply smaller sensitivity relative to the SQL; a geometry in which the nanotube sags close to the gate implies a larger sensitivity but still within the range of Eq. (10).
The uncertainty could be reduced by calibrating the displacement using a measurement of Brownian motion.
Imprecision below the SQL is possible if measurement backaction excites the resonator out of its ground state\cite{Clerk2010}.

In fact, some backaction is evident in Fig.~\ref{fig:3}(b), which shows that the gate voltage range with the strongest signal also leads to a broader mechanical resonance.
Figure~\ref{fig:3}(e) confirms this by plotting the mechanical linewidth against transconductance, showing that higher transconductance, and therefore stronger measurement, correlates with a broader line.
This is because the fluctuating occupation of the SET creates a stochastic force which reduces the mechanical quality factor~\cite{Naik2006, Meerwaldt2012}.
However, the dependence on source-drain bias is opposite (Fig.~\ref{fig:3}(d)).
Both behaviors have been explained by considering the damping mechanism\cite{Clerk2005}.
Electromechanical damping occurs when the changing displacement brings the SET's chemical potential alternately above and below the Fermi level in one of the leads, so that electrons preferentially tunnel onto the SET with low energy and off with high energy.
This damping is reduced by detuning the quantum dot chemical potential from one or both Fermi levels.
In the situation of this measurement, where the bias is large compared to both the thermal energy and lifetime broadening, this detuning is achieved by separating the two Fermi levels with a source-drain bias (Fig.~\ref{fig:3}(d)) or by tuning the SET chemical potential between the Fermi levels with a gate voltage (Fig.~\ref{fig:3}(e)).
This behaviour has been previously observed and modelled quantitatively\cite{Meerwaldt2012}.
At zero bias, occupation fluctuations create mechanical backaction without contributing to the signal.
Increasing the bias therefore increases the efficiency of the measurement, allowing the SQL to be approached more closely.
This description ignores the effect of transport resonances and energy-dependent tunneling, which lead to more complex backaction even including negative damping\cite{Clerk2005}.
A more complete theory of this RF displacement sensor would need to include these effects to assess the potential for quantum-limited measurement.

Finally we consider what limits the achieved displacement imprecision given by Eq.~\eqref{eq:Su}.
The imprecision is approximately as expected given the noise of the cryogenic amplifier (see supplementary material), and could therefore be improved by using an improved superconducting amplifier~\cite{Muck2010}.
Alternatively, the conversion of displacement to current could be improved with larger DC gate voltage, while increasing the quality factor of the tank circuit could give larger transimpedance.
Ultimately, the sensitivity will be limited by the SET's shot noise~\cite{Kafanov2009, Wang2017}.

In conclusion, this experiment shows how to monitor a vibrating carbon nanotube with low noise and high speed using an integrated SET transducer.
Such a device could monitor weak and transient forces on the nanoscale, for example in scanning probe microscopy.
This resonator also approaches the quantum regime both in terms of the ratio of phonon energy to thermal energy (approximately 1:3) and in terms of the measurement sensitivity.
This work therefore opens the way to measuring dynamic electron-phonon coupling effects.

See supplementary material for details on fabrication, the derivation of Eq.~\eqref{eq:Pout}, modelling of the impedance matching circuit, and details of how the uncertainties in Eqs.~(\ref{eq:Su}-\ref{eq:SQLratio}) are calculated.
We acknowledge DSTL, EPSRC (EP/J015067/1, EP/N014995/1), TWCF, FQXI, and the RAEng. We thank S.~Kafanov and A.~Romito for comments and A.W. Robertson for microscopy.

\newpage

\clearpage

	\setcounter{equation}{0}
	\setcounter{figure}{0}
	\setcounter{table}{0}
	\setcounter{page}{1}
	\makeatletter
	\renewcommand{\theequation}{S\arabic{equation}}
	\renewcommand{\thefigure}{S\arabic{figure}}
	\renewcommand{\thetable}{S\Roman{table}}
	\renewcommand{\thesection}{S.\Roman{section}}
	\renewcommand{\bibnumfmt}[1]{[S#1]}
	\renewcommand{\citenumfont}[1]{S#1}

\onecolumngrid
\parbox[c]{\textwidth}{\protect \centering \Large \MakeUppercase{Supplementary Material}}
\rule{\textwidth}{1pt}
\vspace{1cm}
\twocolumngrid

\maketitle

\section{Carbon Nanotube synthesis, transfer, and characterization}
Carbon nanotubes were synthesized using chemical vapour deposition (CVD).
A quartz growth chip, previously patterned with pillars $\sim4~\mu$m high, was coated with PMMA (495A6, spun at 8000 rpm for one minute) thick enough to cover most of the chip while allowing the pillars to protrude.
A catalyst fluid, consisting of Al$_2$O$_3$ nanoparticles, (0.375 mg/ml), Fe(NO$_3$)$_3$\textperiodcentered 9H$_2$O (0.5 mg/ml), and MoO$_2$(acac)$_2$ (0.113 mg/ml) dissolved in methanol, was dropped onto this chip, allowed to dry in air for 1 minute, and then blown away with compressed air~\cite{Pei2015}.
The PMMA layer is then dissolved in boiling acetone for 45 seconds and cold acetone for 90 minutes, leaving catalyst particles only on top of the quartz pillars.

For CVD, the chip was heated to \SI{950}{\degreeCelsius} in a tube furnace while exposed to a H$_2$:Ar (1:2) atmosphere, reducing the $\mathrm{Fe(NO_3)_3}$ to $\mathrm{Fe}$.
The atmosphere was then changed to CH$_4$:H$_2$ (1:9) for 30 minutes, during which nanotubes grow~\cite{Kong1998,Pei2017}.
The nanotubes, some of which span pairs of pillars, are afterwards transferred to the device chip by optical alignment and stamping~\cite{Wu2010,Pei2012,Mavalankar2016}.
The nominal thickness of the contact electrodes was 15~nm Cr + 120~nm Au, and of the gate electrodes 10~nm Cr + 15~nm Au.
To improve electrical contact, we clean the gold surface with oxygen plasma before stamping.

To characterize this growth process, nanotubes grown the same way were transferred to a transmission electron microscopy (TEM) grid. The resulting images showed typical diameter $D=4.5 \pm 1.5$~nm and typical number of walls $\NW=2 \pm 1$.

\section{Calibrating the drive signal}
\label{sec:calibratingdrive}

\begin{figure}[b]
	\includegraphics[width=8.5cm]{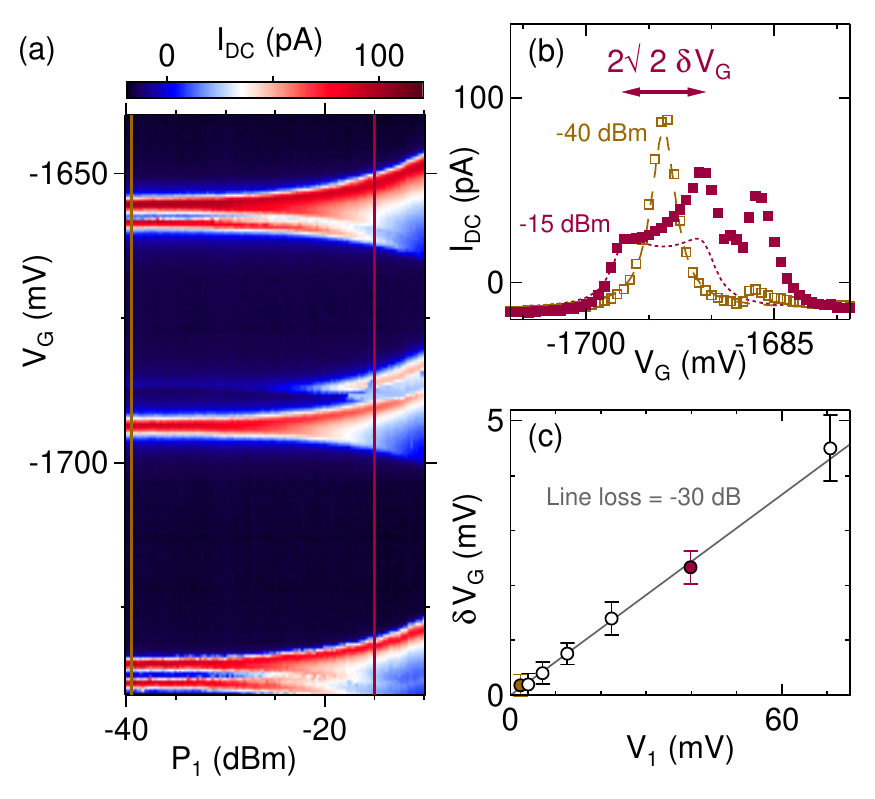}
	\caption{\label{fig:S1} (a) Current through the nanotube as a function of gate voltage and RF drive power into port 1, showing  Coulomb peak broadening at high power. (b) Cuts through the central peak at two different drive powers, with fits (see text). The double arrow marks the peak-to-peak drive voltage with -16~dBm applied, which corresponds to the Coulomb peak splitting. The satellite peaks in (a) and (b) probably correspond to transport through excited states of the dot. (c) Points: Fitted rms amplitude as a function of drive amplitude. Line: Linear fit. Taking account of the impedance mismatch at the gate, the slope gives the attenuation in the cryostat wiring. Throughout this figure, the drive frequency is $f=120$~MHz and the bias is $\VSD=1.6$mV.}
\end{figure}

To calibrate the applied drive amplitude, we measure the broadening of the Coulomb peaks as a function of RF power injected into the cryostat through port 1, with the frequency chosen away from the mechanical resonance~(Fig.~\ref{fig:S1}).
Since the measured DC current is a time average over the RF cycle, this broadening is a measure of the RF amplitude at the gate.
To quantify it, we first fit the middle peak at the lowest applied power assuming lifetime-broadened Coulomb blockade~\cite{Beenakker1991}:
\begin{equation}
\IDC = I_\mathrm{offset} + I_0 \frac{(\GammaV/2)^2}{(\VG-V_0)^2+(\GammaV/2)^2}
\label{eq:unbroadened}
\end{equation}
to extract the intrinsic linewidth $\GammaV = 2.3$~mV.
(Here the fit parameters $I_\mathrm{offset}$, $I_0$, and $V_0$ parameterize the offset current, peak current, and location of the Coulomb peak.)
Keeping all these parameters fixed, we then fit the peak at each power by convolving Eq.~\eqref{eq:unbroadened} with the expected broadening function for a sinusoidal drive:
\begin{equation}
f_\mathrm{P}(\VG) = \frac{1} {\pi \sqrt{2\, \delta \VG^2 - \VG^2} }\, \Theta(2\, \delta \VG^2 - \VG^2)
\end{equation}
where the rms drive amplitude $\delta \VG$ is the fit parameter (Fig.~\ref{fig:S3}(b)) and $\Theta(x)$ is the Heaviside step function.

Plotting the fitted amplitude at the gate against the amplitude $V_1$ applied at port 1 shows that $\delta \VG = 0.061\times V_1$~(Fig.~\ref{fig:S3}(c)).
Taking account of the impedance mismatch at the end of the signal line, which has the effect of doubling the amplitude on the gate,  this implies a line loss of $-30 \pm 1$~dB, as expected from the the nominal $-30$~dB of the attenuators in the cryostat. This measured line loss is the value used to calibrate $\delta \VG$ in the main text.

\section{The electromechanical measurement sensitivity}
In this section, we explain how the response curves of Fig.~3(c) in the main text are fitted, and how the fit parameters lead to an estimate of the displacement sensitivity. In Section~\ref{sec:ResponseFunction} we derive the fit function.
Section~\ref{sec:EstimatingSensitivity} uses the fit parameters to deduce the sensitivity, including error bars.
Section~\ref{sec:ExpectedSensitivity} estimates what sensitivity should be expected given the circuit parameters.

\subsection{The electromechanical response function}
\label{sec:ResponseFunction}
The signal $\Vout$ (see Fig.~1(a) of the main text) arises because the motion excites an oscillating current $I$ through the nanotube and onto the source electrode, and the tank circuit transduces this current to an output voltage $\Vout$.
Here, we show that the corresponding output power is described by Eq.~(1) of the main text.
The output is related to the current by the circuit's transimpedance~\cite{Ares2016}
\begin{equation}
\Ztrans(f) = \Vout/I.
\label{transimpedance}
\end{equation}
With a drive signal $\delta \VG$ at frequency $f$ applied to the gate, the output is therefore
\begin{equation}
V_\mathrm{out}(f) = \Ztrans(f) I + \xi(f) \delta \VG + \VN.
\label{eq:Vout}
\end{equation}
where the first term describes the electromechanical signal, the second term describes the effect of stray signal paths (parameterized by a coupling constant $\xi(f)$), and the third term describes electrical noise. Both $\Ztrans(f)$ and $\xi(f)$ depend on frequency.

The current $I$ contains two terms: a contribution from transport through the nanotube, proportional to the displacement $u$; and a contribution proportional to the velocity $\dot{u}$, arising because the motion modulates the gate capacitance:
\begin{align}
I 		&=	\frac{\partial I}{\partial u}u +  \frac{\partial q}{\partial u} \dot{u}\\
		&= \frac{\partial q}{\partial u} \left( \frac{1}{\CNT} \frac{\partial I}{\partial \VG}u + \dot{u} \right)
		\label{eq:currentsignal}
\end{align}
where $q \equiv \sum_k C_k V_k$ is the charge induced on the quantum dot~\cite{VanHouten2005} by the nearby electrodes. Here the index $k$ labels the electrodes, each set to voltage $V_k$ and coupled to the quantum dot by capacitance $C_k$.
In previous work\cite{Ares2016}, where the tank circuit was connected to a gate electrode, the transport contribution was zero ($\partial I/ \partial \VG = 0$), but here it is the largest part of the signal.

We now assume a linear mechanical response as in Eq.~(1) of the main text, so that the displacement is related to the driving signal by
\begin{equation}
u(f) = \chiM(f) \, \CNT' \VG \, \delta \VG,
\end{equation}
where $\chiM(f) = \frac{1}{4\pi^2m}[\fM^2-f^2 + i f \fM/\QM]^{-1}$ is the mechanical susceptibility, and a prime denotes a derivative
\footnote{Displacement $u$ is defined towards the substrate, so that $C'$ is positive.}
with respect to $u$.
Taking the Fourier transform of Eq.~\eqref{eq:currentsignal} and inserting into Eq.~\eqref{eq:Vout} gives for the output signal:
\begin{equation}
V_\mathrm{out}(f) = \hat{A} \left( \hat{B} + \frac{\fM^2/\QM}{\fM^2 - f^2 + i \frac{f \fM}{\QM}}\right) \delta \VG + \VN
\label{eq:Vout2}
\end{equation}
where we define the complex parameters:
\begin{align}
\hat{A} &\equiv \frac{\Ztrans(\fM) q' \CNT' \VG \QM}{4\pi^2m\fM^2} \left( \frac{1}{\CNT} \frac{\partial I}{\partial \VG} + 2\pi i \fM \right)	\label{eq:defnA}\\
\hat{B} &\equiv \xi(\fM)/\hat{A},	\label{eq:defnB}.
\end{align}
The derivative of the induced charge is
\begin{equation}
q' = \sum_k C_k'V_k,
\label{eq:chargederivative}
\end{equation}
with the sum running over gate electrodes only since the source and drain electrodes are held near zero potential.
Henceforth we take $\hat{A}$ and $\hat{B}$ as constants because the frequency dependence of the electrical circuit is weaker than the frequency dependence of the mechanical response.

The fitting function for Fig.~3(c) of the main text is obtained by converting the voltage of Eq.~\eqref{eq:Vout2} into a detected power:
\begin{align}
\Pout (f)	&= |\Vout(f)|^2/Z_0	\\
&= \frac{A^2}{Z_0} \left| B e^{i\phi_B} + \frac{\fM^2/\QM}{\fM^2 - f^2 + i \frac{f \fM}{\QM}}\right|^2 \delta \VG^2
+ \frac{S_{VV} \Delta f}{Z_0}
\end{align}
where $Z_0 = 50~\Omega$ is the line impedance, $S_{VV}$ is the voltage noise one-sided spectral density, and $\Delta f$ is the resolution bandwidth of the network analyser. We have defined $A \equiv \mathrm{Re}\{\hat{A}\}$, $B \equiv \mathrm{Re}\{\hat{B}\}$, and $\phi_\mathrm{B} \equiv \mathrm{arg}\{\hat{B}\}$.
This is Eq.~(1) of the main text, with fit parameters $\fM$, $\QM$, $A$, $B$, $\phi_M$, and $S_{VV}$.

\begin{table*}\begin{tabular}{ |l c|c c c|c|l|}
		\hline\hline
		\multicolumn{2}{|c|}{Parameter}		& \multicolumn{3}{c|}{Estimate}			& Unit 			& Method\\
									&		& High imprecision 		& Best		 	& Low imprecision		&  				& \\
		\hline
		\hline
		Mechanical frequency		& $\fM$	& 179.50		& 179.53 		& 179.56		& MHz 			&
		\multirow{4}{*}{Fit Fig.~3(c)}\\
		Quality factor				& $\QM$	& 7500		& 6000			& 4500		& -				& \\
		Electromechanical signal	& $\VM$	& 8.7		& 8.8 			& 	8.9		& nV			& \\
		Voltage noise				& $\sqrt{S_{VV}}$	& 50.8			& 50.3				& 49.8			& $\mathrm{pV/\sqrt{Hz}}$	& \\
		\hline

		Nanotube diameter			& $D$	& 3			& 4.5 			& 6			& nm 			&
		\multirow{2}{*}{TEM}\\
		Number of walls				& $\NW$	& 1			& 2 			& 3			& -		 		& \\
		\hline

		Suspended height			& $H$				& 4.5			& 18			& 110		& nm 			&
		\multirow{4}{*}{Electrostatic simulation}\\
		Misalignment					& $\theta$ 			& 0		& 35			& 70		& $^\circ$			& \\
		Effective charge derivative		& $\partial{q}/\partial{u}$
														& -344		& -53.2			& -14.8		&  pC/m 			& \\
		Capacitance derivative		& $\partial{\CNT}/\partial{u}$
														& 320		& 49.4			& 13.1		&  pF/m 			& \\
		\hline

		Mass						& $m$				& $5.8$		& $21.3$			& $102$		& ag 			&
		\multirow{1}{*}{Eq.~\eqref{eq:mass}}\\
		\hline

		Excitation voltage			& $\delta \VG$		& $1.6$			& $1.4$				& $1.3$			& $\mu$V		&
		\multirow{1}{*}{Section \ref{sec:calibratingdrive}}\\
		\hline

		RMS displacement			& $u$				& 836		& 25.2			& 0.9		& pm 			&
		\multirow{1}{*}{Eq.~\eqref{eq:uvalue}}\\
		\hline

		Displacement sensitivity	& $\sqrt{S_{uu}}$				& 4830		& 144.3			& 5.3		& $\mathrm{fm/\sqrt{Hz}}$ &  Main text Eq.~(6)\\
		Expected sensitivity	& $\sqrt{S_{uu}^\mathrm{expected}}$				& 106		& 574			& 2596		& $\mathrm{fm/\sqrt{Hz}}$ & Eq.~\eqref{eq:SuExpected}\\
		Standard quantum limit	& $\sqrt{S_{uu}(\mathrm{SQL})}$				& 10.3		& 4.84			& 1.91		& $\mathrm{fm/\sqrt{Hz}}$ & Main text Eq.~(8)\\
		\hline\hline
	\end{tabular}
	\caption{\label{tab:parameters} Summary of parameters used to calculate the displacement sensitivity, showing the best estimate of parameters, and two  cases corresponding to the highest and the lowest values of the imprecision consistent with the experimental results. The error bar in Eq. (10) of the main text is taken by considering these two extremes.
	}
\end{table*}

\subsection{Estimating the displacement imprecision}
\label{sec:EstimatingSensitivity}

\begin{figure}
	\includegraphics[width=3.37in]{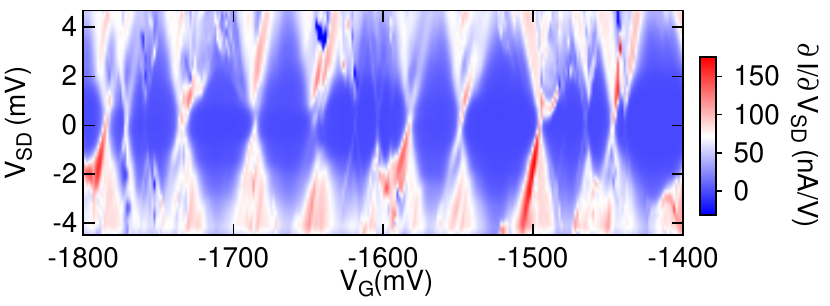}
\caption{\label{fig:S2} Differential conductance through the nanotube as a function of the voltage $\VG$ on gate 2 and of the bias $\VSD$ across the quantum dot. The width of the Coulomb diamonds is $\Delta \VG = 49 \pm 8$~mV, from which the corresponding gate capacitance $\CNT=3.3 \pm 0.6$~aF is extracted.}
\end{figure}

To calculate the displacement imprecision, we must estimate the parameters appearing in Eq.~(5) of the main text, repeated here:
\begin{equation}
u = \frac{1}{4\pi^2 m} \left|\frac{1}{\fM^2-f^2+i\frac{f \fM}{\QM}}\right| \VG  \CNT' \, \delta \VG.
\label{eq:urms}
\end{equation}
For the data of Fig.~3(c), we estimate these as follows:
\begin{itemize}
	\item The parameters $\fM=179.53 \pm 0.03$~MHz and $\QM = 6000 \pm 1500$ are deduced by fitting Fig.~3(c) of the main text.
	\item The gate capacitance derivative $\CNT'$, which sets the electrostatic force, is estimated from a finite-element electrostatic simulation.
Setting up the simulation is made difficult by the fact that the height $H$ of the nanotube above the gate, the nanotube's misalignment $\theta$, and its diameter~$D$ are not known precisely (main text Fig.~1(a)).
However, these quantities are constrained by the fact that the gate capacitance $\CNT$ is known.
From the Coulomb peak spacing for the SET (Fig.~\ref{fig:S2}) the gate capacitance is $\CNT = e/\Delta \VG = 3.3 \pm 0.6$~aF.
This lets us use the simulation to determine the minimum and maximum values of $\partial \CNT/{\partial u}$ consistent with this value of $\CNT$ and with the geometric constraints.
We expect from TEM images that $D=4.5 \pm 1.5$~nm, while the range of $H$ is set by the thickness of the contact electrodes, giving $H<110$~nm.

The maximum misalignment consistent with these constraints is $\theta=70^\circ$, which can occur when $D=6$~nm and $H=110$~nm.
If there is no misalignment ($\theta=0^\circ$) and the diameter is $D=3$~nm, the suspended height is required to be as small as $H=4.5$~nm.
As the preferred estimate for $\theta$ we take the central value, giving $\theta=35\pm35$ degrees and implying $H=18_{-13.5}^{+92}$~nm and $\partial \CNT/{\partial u} = 49.4^{+271}_{-40.5}$~pF~m$^{-1}$. The numerical derivative is calculated by simulating a uniform displacement of the nanotube, and multiplying by $\sqrt{2}$ to take account of the fact that the actual mode shape has a larger displacement near the centre~\cite{Poot2012}.

\item The nanotube's mass is given by:
\begin{equation}
m 	= \frac{\pi \rho_S \NW D l}{\cos \theta} = 21.3^{+81}_{-15.5}~\mathrm{ag}
\label{eq:mass}
\end{equation}
where $l=800$~nm is the contact spacing, $\rho_S=7.7 \times 10^{-7}~\mathrm{kg~m}^{-2}$ is the sheet density of graphene and $\NW=2\pm 1$ is the number of nanotube walls estimated by TEM.

\item The drive amplitude $\delta \VG$ is known from the RF transmission of the cryostat wiring, and confirmed by measuring Coulomb peak broadening~(Section~\ref{sec:calibratingdrive}).

\end{itemize}

Substituting these values into Eq.~\eqref{eq:urms} leads to an estimate for the rms displacement on resonance:
\begin{equation}
u=25.2^{+811}_{-24.3}~\mathrm{pm}
\label{eq:uvalue}
\end{equation}
and therefore a displacement sensitivity, defined as the square root of the measurement imprecision~\cite{Clerk2010}:
\begin{align}
\sqrt{S_{uu}}	&= \frac{u}{\VM}\sqrt{S_{VV}}								\label{eq:Suformula}\\
	&= 144.3^{+4686}_{-139}~\mathrm{fm/\sqrt{Hz}}.		\label{eq:Suvalue}
\end{align}
These estimated parameters and their contributions to the uncertainty of $S_{uu}$ are summarised in Table~\ref{tab:parameters}.

\subsection{The expected displacement sensitivity}
\label{sec:ExpectedSensitivity}
We now compare this experimentally extracted displacement sensitivity (Eq.~\eqref{eq:Suvalue}) with the value expected from the electrical characteristics of the circuit.
On resonance, the electromechanical signal is given by the first term of Eq.~\eqref{eq:Vout}:
\begin{align}
\VM 	&= \Ztrans(\fM) I	\\
		&= \Ztrans(\fM) \frac{q'}{\CNT} \frac{\partial \IDC}{\partial \VG}u,
\end{align}
where $I$ is taken from Eq.~\eqref{eq:currentsignal}, neglecting the small second term.
This implies a sensitivity
\begin{align}
\sqrt{S^\mathrm{expected}_{uu}} 	&= \frac{u}{\VM} \sqrt{S_{VV}}	\\
	&= \frac{1}{\Ztrans(\fM)} \frac{1}{q'} \frac{\CNT}{ \frac{\partial \IDC}{\partial \VG}} \sqrt{S_{VV}}.	\label{eq:SuExpected}
\end{align}
The transimpedance of the circuit is not directly measurable but can be plausibly modelled (see Section \ref{sec:modelTank}), leading to $\Ztrans(\fM) \approx 198~\Omega$. Taking the slope $\frac{\partial \IDC}{\partial \VG} \approx 28~\mathrm{nA/V}$ from data similar to Fig.~3(a) of the main text,
taking the charge derivative $q' \approx -50~\mathrm{pC/m}$ from Eq.~\eqref{eq:chargederivative} using electrostatic simulation of the gate capacitances,
and taking the noise $\sqrt{S_{VV}} \approx 51~\mathrm{pV}/\sqrt{\mathrm{Hz}}$ from the specification of the cryogenic amplifier, leads by Eq.~\eqref{eq:SuExpected} to an expected sensitivity:
\begin{equation}
\sqrt{S^\mathrm{expected}_{uu}} \sim 600~\mathrm{fm} / \sqrt{\mathrm{Hz}}.
\label{eq:Suexpected}
\end{equation}

This is within the uncertainty of the measurement (Eq.~\eqref{eq:Suvalue}), implying that the motion is approximately as estimated in Eq.~\eqref{eq:uvalue}, that the electrical signal indeed arises from changes in the nanotube's conductance, and that the noise is dominated by the electrical noise of the cryogenic amplifier.
However, we note that this agreement depends on the estimate of $\Ztrans$, and that including parasitic impedances in our circuit model can lead to poorer agreement between Eq.~\eqref{eq:Suvalue} and Eq.~\eqref{eq:Suexpected}, possibly implying a larger nanotube mass than stated in Eq.~\eqref{eq:mass}.

\section{Modeling the matching circuit}
\label{sec:modelTank}
\begin{figure}
	\includegraphics[width=3.37in]{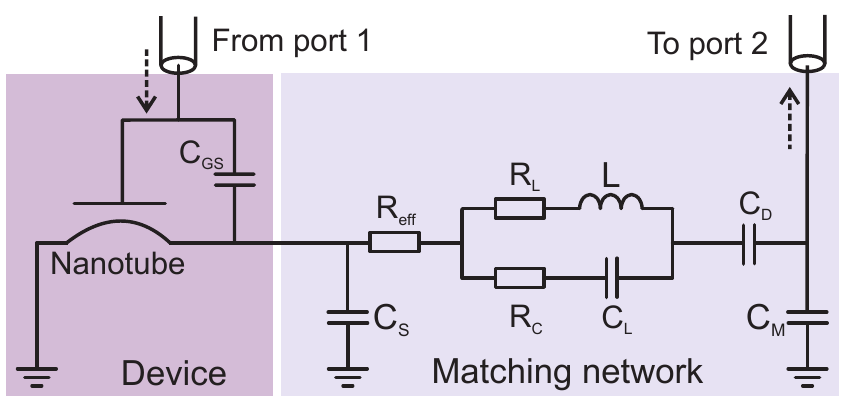}
\caption{\label{fig:S3} Lumped element model of the impedance matching tank circuit. For an explanation of component values, see text. The nanotube has a large resistance, and is therefore not included in the model.}
\end{figure}

This section explains how the fits in Fig.~2(a) of the main text are made, and how these lead to the estimate of $\Ztrans$ used in Eq.~\eqref{eq:SuExpected}. The impedance matching tank circuit is modelled as shown in Fig.~\ref{fig:S3}. The circuit is constructed from an on-board inductor $L$, fixed capacitors $C_\mathrm{D}=87$~pF and $C_\mathrm{M}$, and a tunable capacitor $\CS$. In practice, the circuit also incorporates significant parasitic impedances, which we take account of as follows~\cite{Ares2016}. The self-resonance and loss of the inductor (Coilcraft 1206CS-221) are modelled by adding a capacitor and resistors as shown, with fixed values given by the component datasheet: $L=223$~nH, $C_\mathrm{L} = 0.082$~pF, $R_\mathrm{C} = 25~\Omega$, and $R_\mathrm{L}= K\times \sqrt{f_\mathrm{C}}$ with $K=\SI{3.15e-4}{\Omega/\sqrt{Hz}}$. Dissipation elsewhere in the circuit is modelled by an effective resistance $R_\mathrm{eff}$. The parasitic coupling between the driving gate and the source electrode, which excites the tank circuit independently of the nanotube's motion, is modelled by a capacitor $C_\mathrm{GS}$.

The values of the unknown parasitics were estimated by fitting the transmission traces in Fig.~2(a) of the main text. The trace with $\VS=0$~V is fitted with the following four free parameters: $C_\mathrm{S}=\SI{3.82}{pF}$, $R_\mathrm{eff}=\SI{5.04}{\Omega}$, $C_\mathrm{GS}=\SI{14.5}{fF}$, and $C_\mathrm{M}=\SI{80}{pF}$.
Although it is possible that other specified component values change substantially at low temperature, in fact these four free parameters lead to good agreement with the data.
For the other traces, only the behaviour of the varactor should change, and the fits therefore hold $C_\mathrm{GS}$ and $C_\mathrm{M}$ fixed, obtaining fair agreement with the data using only $C_\mathrm{S}$ and $R_\mathrm{eff}$ as free parameters.
With all parameters in the model circuit of Fig.~\ref{fig:S3} now estimated, the value of the transimpedance $\Ztrans$ then follows from Kirchoff's laws.

\section*{references}

\vfill
\end{document}